\documentclass[conference]{IEEEtran}

\usepackage{amsmath}
\usepackage{amsfonts}
\usepackage{amssymb}
\usepackage{amsmath}
\usepackage{cite}
\usepackage{subfig}
\usepackage{graphicx}
\usepackage{algorithm,algcompatible}
\algnewcommand\INPUT{\item[\textbf{Input:}]}%
\algnewcommand\OUTPUT{\item[\textbf{Output:}]}%

\begin{document}

\title{Demonstration of a 1.2 Gbps Always-on Fully-Connected Mesh Network with RFSoC SDRs

\thanks{This work was supported by NSF Grants CNS-2117822 and EEC-2133516 and AFRL Grant FA8750-20-C-1021. Distribution A. Approved for public release: Distribution Unlimited: AFRL-2026-0915 on 2 Mar 2026.}}  

\IEEEoverridecommandlockouts

\author{\IEEEauthorblockN{ 
    Hatef Nouri$^*$, George Sklivanitis$^*$, Dimitris A. Pados$^*$, and Elizabeth Serena Bentley$^{\dagger}$}
\IEEEauthorblockA{$^*${Center for Connected Autonomy and AI, Florida Atlantic University,} Boca Raton, FL 33431 USA\\
$^{\dagger}$Air Force Research Laboratory, Rome, NY, 13441, USA\\
\{hnouri, gsklivanitis, dpados\}@fau.edu,  elizabeth.bentley.3@us.af.mil}
}

\maketitle

\begin{abstract}
We design and implement on Radio Frequency System-on-Chip (RFSoC) software-defined radios (SDRs) a complete-graph network of four unmanned aerial vehicles and demonstrate real-time 4K video streaming over twelve always-on 2x2 multiple-input multiple-output (MIMO) links. The testbed operates at an aggregate network throughput of $\approx$1.2 Gbps (i.e., 12 links of 99.84 Mbps) across a shared bandwidth of 200 MHz. To the best of our knowledge, this is the first demonstration of low-latency digitally controlled frequency-division duplex (FDD) RFSoC-based MIMO wireless links capable of  simultaneously supporting multiple real-time, uncompressed 4K video streams.
The testbed consists of four AMD/Xilinx Zynq UltraScale+ RFSoC ZCU111 evaluation kits configured as a fully-connected mesh network with custom-built physical and medium-access-control layers, adaptive equalization, and adjacent-band filtering implemented entirely in RFSoC's programmable logic. A host-side graphical user interface (GUI) provides real-time visualization of each link's performance including error vector magnitude (EVM), pre-detection signal-to-interference-plus-noise ratio (SINR), and bit error rate (BER), and enables dynamic reconfiguration of link parameters during operation.
\end{abstract}


\IEEEpeerreviewmaketitle

\section{Introduction}

Networked unmanned aerial vehicle (UAV) systems have emerged as a promising solution for applications such as surveillance, search-and-rescue, coordinated mapping, and distributed sensing. Resilient, high-throughput connectivity is necessary for enabling cooperative behavior, synchronized operation, and real-time data exchange among UAVs \cite{UAV1,UAV2}.

\begin{figure}[t]
\centering
\includegraphics[width=0.7\columnwidth]{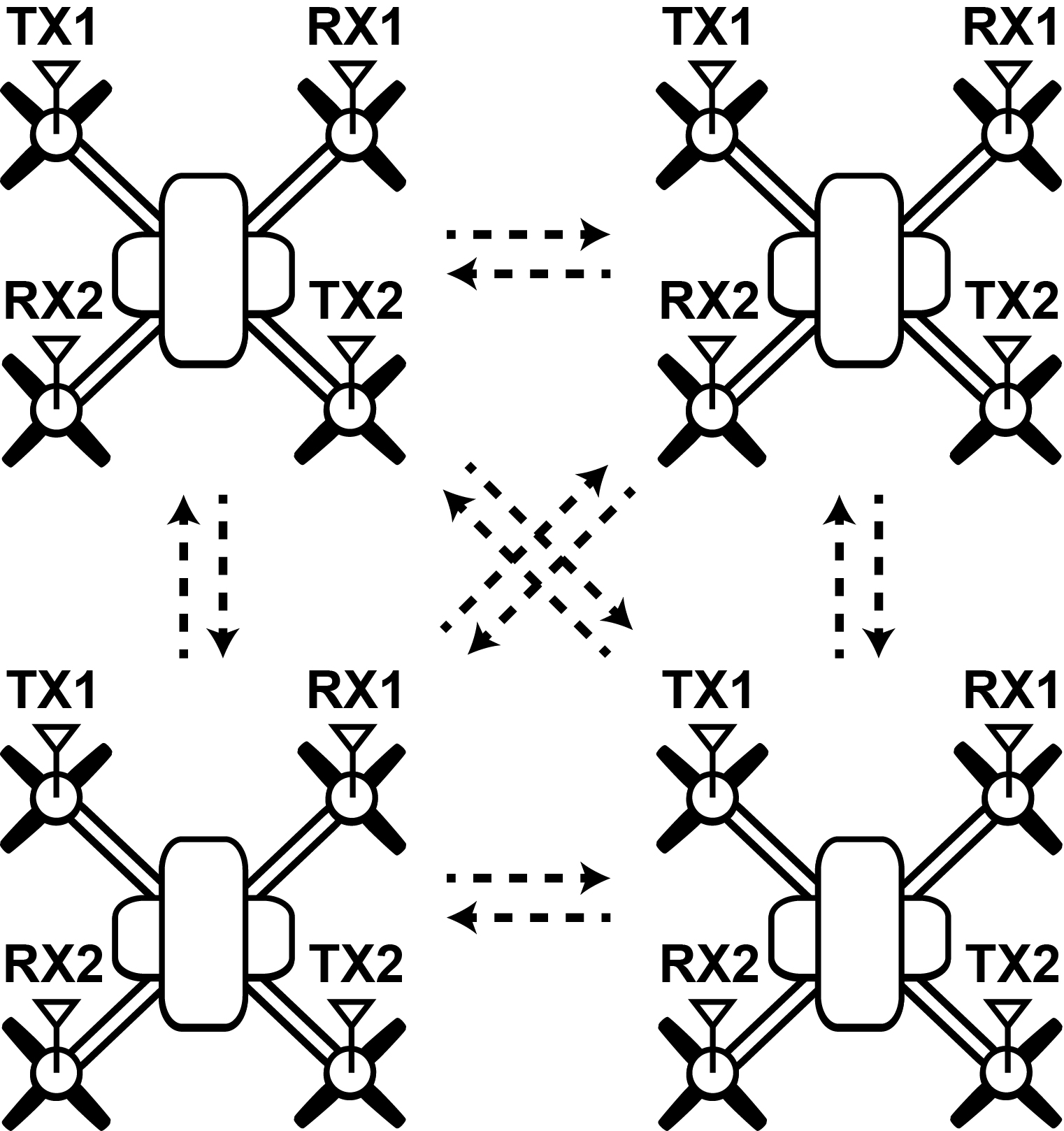}
\caption{Complete-graph network of four UAVs.}
\label{OTA}
\end{figure}

\begin{figure}
\centering
\includegraphics[width=0.8\columnwidth]{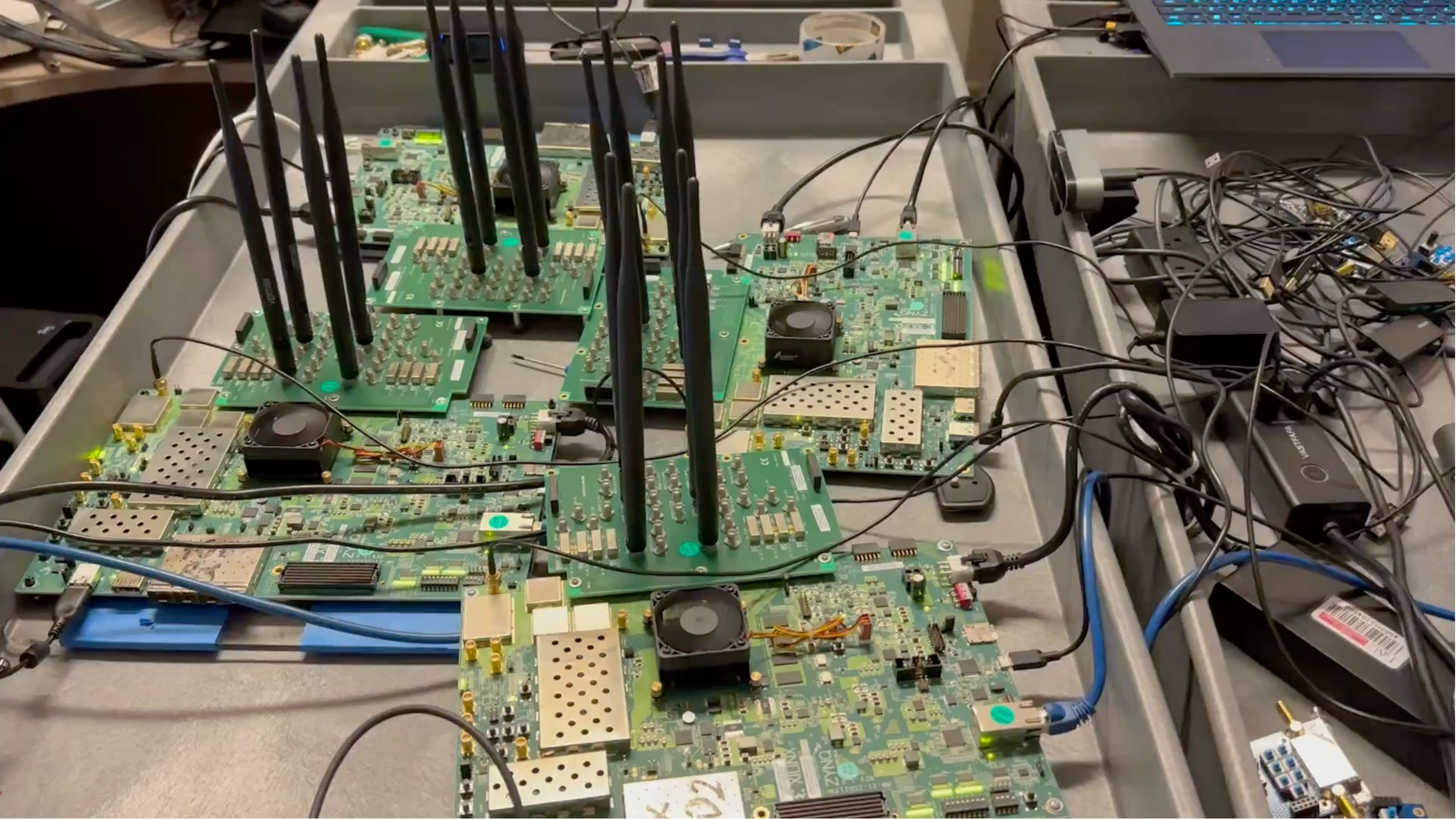}
\caption{Over-the-air (OTA) software-defined radio testbed with 2×2 MIMO RFSoC SDRs.}
\label{RFSoCTestbed}
\end{figure}

We design and develop a fully connected (complete graph) network of four UAV nodes, each equipped with two transmit and two receive antennas and the AMD/Xilinx Zynq UltraScale+ RFSoC ZCU111 software-defined radio (SDR) module. The UAVs are arranged in a cross (+) configuration, which provides spatial diversity and reduces shadowing effects to sustain high-throughput operation (Fig.~\ref{OTA}). Medium-access control is based on frequency-division multiple access (FDMA) principles where adjacent frequency bands are distributed to each UAV transmitter \cite{UAV3} to mitigate self-interference and support simultaneous (full duplex) always-on operation. The RFSoC architecture enables each one of the two receiver antennas to directly sample the transmitted RF signals from all UAVs and implement fully digital downconversion and low-pass filtering, thus reducing hardware compexity and cost. Figure \ref{RFSoCTestbed} depicts the SDR testbed with four 2x2 MIMO RFSoCs, replicating the fully connected network of UAVs. 

The AMD/Xilinx Zynq UltraScale+ RFSoC ZCU111 SDR platform integrates eight 12-bit 4.096~GSps analog-to-digital converters (ADCs) and eight 14-bit 6.554~GSps digital-to-analog converters (DACs). The RFSoC heterogeneous architecture facilitates efficient hardware–software co-design \cite{drozdenko2017hardware,jiao2020openwifi,warp} in which physical-layer baseband processing including modulation, MIMO combining, filtering, and equalization is implemented in the programmable logic (PL) to achieve deterministic low-latency operation. The ARM-based processing system (PS) manages system control, I/O handling, real-time monitoring of link parameters, and user-side configuration through a host-based graphical interface.

\begin{figure*}[t]
\centering
\includegraphics[width=0.9\textwidth]{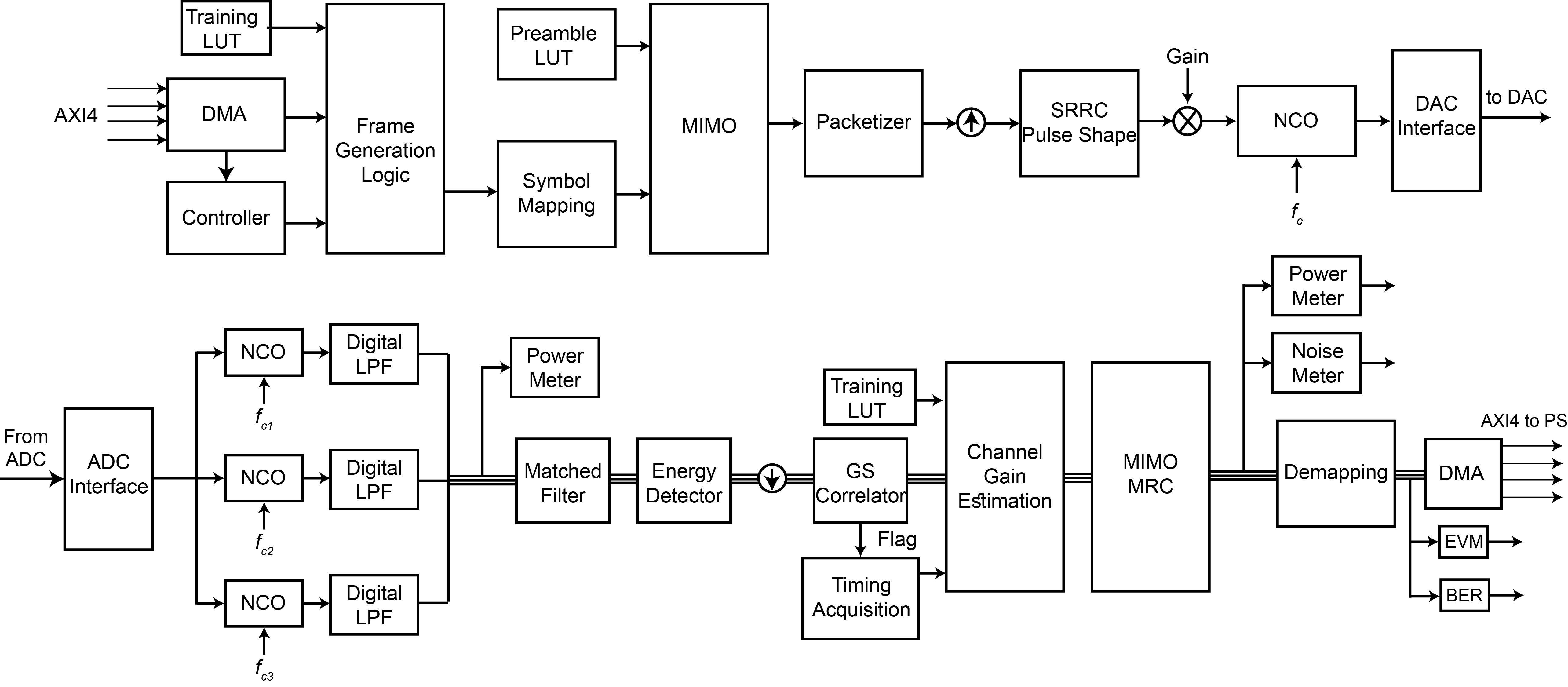}
\caption{High-level hardware architecture of wireless transmitter and receiver.}
\label{fig:txrxarch}
\end{figure*}

In this demonstration, we show a fully connected network of four 2x2 MIMO RFSoC-based nodes, i.e., twelve always-on MIMO links. The system is designed to support 99.84~Mbps per link using 16-QAM modulation over a 37.44~MHz bandwidth, enabling simultaneous all-to-all data exchange across UAVs. A key challenge addressed in this work is the transition from a simulation-based model in Simulink to a hardware architecture operating at 200~MHz FPGA clock. This required extensive re-design of the ADC/DAC interface, digital frequency translation using numerically controlled oscillators (NCOs), designing and implementing low-pass filters with steep slope (roll-off) for adjacent band isolation, and optimized equalizer architectures capable of meeting stringent timing constraints. Specifically, we implement cascaded high-order FIR filters, and adaptive symbol-spaced equalization following maximum-ratio combining (MRC) to mitigate inter-symbol interference (ISI) and channel selectivity under mobility. To enable decoding of each UAV data stream, we implement AXI4-Stream data switching. Real-time PS/PL control enables up to twelve concurrent OTA communication links. 
Experiments demonstrate stable full-duplex operation with BER below 1e-5 and SNR of 28–30~dB across all links. At the application layer, we choose to transmit raw/uncompressed 4K video from each UAV to demonstrate the end-to-end throughput and low-latency capabilities of the network.

\section{Transmit and Receive Processing Architecture}


Each RFSoC node is interfaced with an XM500 RFMC balun card and a custom RF front end operating in the 900~MHz band, consisting of antennas, amplifiers, and filters optimized for low loss and efficient power delivery. The RFSoC ADCs and DACs operate at 3.93216~GSps, enabling operation in the first Nyquist zone. Digital decimation and interpolation stages reduce the sampling rates to FPGA-compatible clock domains. Each frame consists of a fixed 64-byte preamble followed by a PHY data frame containing MIMO training, pilots, header, payload, and CRC.

\textbf{Transmitter Design:} The transmitter architecture is shown in Fig.~\ref{fig:txrxarch}. Data is streamed from the PS to the PL via AXI4-Stream DMA and buffered in FIFOs. A controller orchestrates frame assembly, while LUT-based blocks insert preambles and training sequences. Payload bits are mapped to Gray-coded 16-QAM symbols and organized into a 2x2 MIMO structure, where each transmit antenna is activated in a dedicated interval for channel estimation. A programmable gain stage controls transmit energy. Pulse shaping is performed using a 65-tap square-root raised cosine (SRRC) interpolation filter, followed by digital upconversion using an NCO before transmission through the DACs.

\textbf{Receiver Design:} At the receiver side, each one of the two ADC channels captures and samples the RF signals at 3.93216~GSps and decimate them to baseband sample rates. Digital downconversion is performed using NCO-based mixers. A bank of three low-pass filters is then used to isolate the frequency band of each UAV transmitter. A matched SRRC filter maximizes SNR prior to synchronization. Frame detection relies on correlation with a known Golay-based preamble, using adaptive thresholding to maintain robustness under fading and interference.

Symbol timing is recovered using a moving-average timing metric. Channel estimation is performed using the known MIMO training sequences, and the dominant channel tap is used for normalization. For the 2x2 MIMO configuration, MRC coherently combines the two receive streams. Demodulation, bit decoding, and data reformatting are implemented entirely in hardware and are streamed back to the PS via AXI4-Stream DMA. 

Real-time performance monitoring is integrated into hardware. BER, EVM, and pre-detection SINR are computed using gated power measurements over training and payload symbol intervals, enabling continuous link-quality tracking.

\bibliographystyle{IEEEtran}
\bibliography{sample}
%

\end{document}